\documentstyle[12pt]{article}

\newcommand{\q}{\underline{q}}
\newcommand{\sq}{\scriptstyle{\underline{q}}}

\begin{document}

\title{
Even and odd $q$-deformed charge coherent states and their
nonclassical properties}
\author{{X.-M. Liu$~^{{\tt a,b}}$, C. Quesne$~^{{\tt b,*}}$} \\
{\small ${}^{{\tt a}}$ Department of Physics, Beijing Normal University,
Beijing 100875, China }\\
{\small ${}^{{\tt b}}$ Physique Nucl\'{e}aire Th\'{e}orique et Physique
Math\'{e}matique,}\\
{\small  Universit\'{e} Libre de Bruxelles, Campus de la Plaine CP229,}\\
{\small Boulevard du Triomphe, B-1050 Brussels, Belgium ${}^1$}}
\date{}
\maketitle


\begin{center}
\begin{minipage}{120mm}
\vskip 0.15in
\baselineskip 0.374in
\begin{center}{\bf Abstract}\end{center}
\mbox{}\hspace{6mm}Even and odd $q$-deformed charge coherent states are
constructed, their (over)completeness proved and their generation explored. A
$D$-algebra realization of the SU$_q$(1,1) generators is given in terms of them. They
are shown to exhibit SU$_q$(1,1) squeezing and two-mode $q$-antibunching, but
neither one-mode, nor two-mode $q$-squeezing.

PACS number(s): 42.50.Dv; 03.65.-W; 03.65.Ca; 03.65.Fd
\end{minipage}
\end{center}

\vskip 1mm 

{\small ${}^*$ Corresponding author}

{\small ~~E-mail addresses: liuxm@263.net (X.-M. Liu),
cquesne@ulb.ac.be (C. Quesne).}

{\small ${}^1$ Mailing address} 


\newpage

\baselineskip 0.374in

\section*{1. Introduction}

\mbox{}\hspace{6mm}The coherent states introduced by Schr\"{o}dinger [1] and
Glauber [2] are the eigenstates of the boson annihilation operator, and have
widespread applications in the fields of physics [3$-$7]. However, in all
the cases the quanta involved are uncharged. In 1976, Bhaumik {\sl et al.}
[4,8] constructed the boson coherent states which, carrying definite charge,
are the eigenstates of both the pair boson annihilation operator and the
charge operator. This kind of states are the so-called charge coherent
states. Based on this work, the charge coherent states for SU(2) [9,10], 
SU(3) [11], and arbitrary compact Lie groups [12] were also put forward.

The concept of charge coherent states has proved to be very useful in many
areas, such as elementary particle physics [13$-$17], quantum field theory
[12,18,19], nuclear physics [20], thermodynamics [21$-$23], quantum
mechanics [24], and quantum optics [25$-$27]. Moreover, some schemes for
generating charge coherent states in quantum optics were proposed
[25,26,28,29].

As is well known, the even and odd coherent states [30], which are the two
orthonormalized eigenstates of the square of the boson annihilation
operator, play an important role in quantum optics [31$-$33]. In a previous
Letter [34], inspired by the above idea, one of the authors (X.-M.L.) has
generalized the charge coherent states to the even and odd charge coherent
states, defined as the two orthonormalized eigenstates of both the square of
the pair boson annihilation operator and the charge operator.

On the other hand, quantum groups [35,36], introduced as a mathematical
description of deformed Lie algebras, have given the possibility of
generalizing the notion of coherent states to the case of $q$-deformations
[37$-$41]. A $q$-deformed harmonic oscillator [37,42] was defined in terms
of $q$-boson annihilation and creation operators, the latter satisfying the
quantum Heisenberg-Weyl algebra [37,42,43], which plays an important role in
quantum groups. The $q$-deformed coherent states introduced by Biedenharn
[37] are the eigenstates of the $q$-boson annihilation operator. Such states
have been well studied [38,39,44,45], and widely applied to quantum optics
and mathematical physics [41,46$-$50]. Furthermore, the $q$-deformed charge
coherent states [51,52] were constructed as the eigenstates of both the
pair $q$-boson annihilation operator and the charge operator.

A natural extension of the $q$-deformed coherent states is provided by the
even and odd $q$-deformed coherent states [53], which are the two
orthonormalized eigenstates of the square of the $q$-boson annihilation
operator. In a parallel way, it is very desirable to generalize the $q$%
-deformed charge coherent states to the even and odd ones, defined as
the two orthonormalized eigenstates of both the square of the pair $q$-boson annihilation
operator and the charge operator.

This Letter is arranged as follows. In Section 2, the even and odd 
$q$-deformed charge coherent states are constructed. Their completeness is
proved in Section 3. Section 4 is devoted to generating them. In Section 5,
they are used to provide a $D$-algebra realization of the SU$_q(1,1)$
generators. Their nonclassical properties, such as SU$_q(1,1)$ squeezing,
single- or two-mode $q$-squeezing, and two-mode
$q$-antibunching, are studied in Section 6. Section 7 contains a summary of 
the results.


\section*{2. \boldmath Even and odd $q$-deformed charge coherent states}

\mbox{}\hspace{6mm}Two mutually commuting $q$-deformed harmonic
oscillators are defined in terms of two pairs of independent $q$-boson
annihilation and creation operators $a_i$, $a_i^{+}$ $(i=1,2)$, together
with corresponding number operators $N_i$, satisfying the quantum
Heisenberg-Weyl algebra 
\begin{equation}  \label{e1}
{a_i}a_i^{+}-qa_i^{+}{a_i}=q^{-N_i},
\end{equation}
\begin{equation}  \label{e2}
[{N_i},~a_i^{+}]=a_i^{+},\qquad[{N_i},~{a_i}]=-{a_i},
\end{equation}
where $q$ is a positive real deformation parameter. 
The operators $a_i$, $a_i^{+}$, and $N_i$ act in the Fock space with basis $
|n\rangle_i$ $(n=0,1,2,\ldots )$, such that 
\begin{equation}  \label{e3}
{a_i}|0{\rangle}_i=0,\qquad |n{\rangle}_i={\frac{{(a_i^{+})}^n}{%
\sqrt{[n]!}}} |0{\rangle}_i,
\end{equation}
where 
\begin{equation}  \label{e4}
[n]!{\equiv}[n]_q!=[n]_q[n-1]_q\ldots [1]_q,\qquad [0]!=1,
\end{equation}
\begin{equation}  \label{e5}
[n]_q=\frac{q^n-q^{-n}}{q-q^{-1}}{\equiv}[n].
\end{equation}
Their action on the basis states is given by 
\begin{equation}  \label{e6}
{a_i}|n{\rangle}_i=\sqrt{[n]}|n-1{\rangle} _i,\qquad {a_i^{+}}|n{%
\rangle}_i={\sqrt{[n+1]}}|n+1{\rangle} _i,\qquad {N_i}|n{\rangle}%
_i=n|n{\rangle}_i.
\end{equation}
Note that $[n]$ is invariant under $q~{\leftrightarrow}~1/q$. We may therefore choose
$0<q\le 1$. In the following, $[n]$ will refer to the $q$-deformed  $n$ defined by
(\ref{e5}) corresponding to the base $q$. If the base is different, then it will be
indicated explicitly.
 
The $q$-boson operators $a_i$ and $a_i^{+}$ can be constructed from the
conventional boson annihilation and creation operators $b_i$, $b_i^{+}$ in
the following way [54]: 
\begin{equation}  \label{e7}
{a_i}=\sqrt{\frac{[N_i+1]}{N_i+1}}b_i,\qquad {a_i^+}={b_i^+}\sqrt{%
\frac{[N_i+1]}{N_i+1}},
\end{equation}
where $N_i=b_i^{+}b_i$. It is worth noticing that $[N_i]=a_i^{+}a_i$.

We first briefly review the $q$-deformed charge coherent states. The
operators $a_1(a_1^{+})$ and $a_2(a_2^{+})$ are assigned the ``charge''
quanta 1 and $-1$, respectively. Thus the charge operator is given by 
\begin{equation}
Q=N_1-N_2.  \label{e8}
\end{equation}
In view of the fact that 
\begin{equation}
\lbrack Q,a_1a_2]=0,  \label{e9}
\end{equation}
the $q$-deformed charge coherent states are defined as the eigenstates of
both the pair $q$-boson annihilation operator $a_1a_2$ and the charge
operator $Q$, i.e., 
\begin{equation}
a_1a_2|{\xi },\q{\rangle }={\xi }|{\xi },\q{\rangle }, \qquad
Q|{\xi },\q{\rangle }=\q |{\xi },\q{\rangle },  \label{e10}
\end{equation}
where $\xi $ is a complex number and $\q$ is the charge number, which is
a fixed integer. With the help of the two-mode Fock space's completeness
relation 
\begin{equation}
\sum\limits_{m=0}^\infty \sum\limits_{n=0}^\infty |m,n{\rangle }{\langle }%
m,n|=I, \qquad |m,n{\rangle } \equiv |m{\rangle }_1|n{\rangle }_2,
\label{e11}
\end{equation}
where $|m{\rangle }_1$ and $|n{\rangle }_2$ are the eigenstates of $N_1$
and $N_2$ corresponding to the eigenvalues $m$ and $n$, respectively, the
$q$-deformed charge coherent states can be expanded as
\begin{eqnarray}
|{\xi },\q{\rangle }&=&N_{\sq}{\ \sum_{p=\max(0,-\sq)}^\infty }\frac{{%
\xi }^{p+\min(0,\sq)}}{{\{[p]![p+\q]!\}}^{1/2}}|p+\q,p{\rangle }
\nonumber \\ 
&=&\left\{ 
\begin{array}{ll}
N_{\sq}{\sum\limits_{n=0}^\infty }\frac{{\xi }^n}{{\{[n]![n+\sq]!\}}
^{1/2}}|n+\q,n{\rangle }, &\q\geq 0, \\[0.2cm] 
N_{\sq}{\sum\limits_{n=0}^\infty }\frac{{\xi }^n}{{\{[n]![n-\sq]!\}}
^{1/2}}|n,n-\q{\rangle }, & \q\leq 0,
\end{array}
\right.  \label{e12}
\end{eqnarray}
where the normalization factor $N_{\sq}$ is given by 
\begin{equation}
N_{\sq}{\equiv }N_{\sq}({|\xi |}^2)={\ \left\{ \sum\limits_{n=0}^\infty 
\frac{{({|\xi |}^2)}^n}{[n]![n+|\q|]!}\right\} }^{-{1/2}}.
\label{e13}
\end{equation}

We now seek the even and odd $q$-deformed charge coherent states, which are
the two orthonormalized eigenstates of both the square ${(a_1a_2)}^2$ and
the operator $Q $, i.e., 
\begin{equation}
{(a_1a_2)}^2|{\xi },\q{\rangle }_{e(o)}={\xi }^2|{\xi },\q{\rangle }%
_{e(o)}, \qquad Q|{\xi },\q{\rangle }_{e(o)}=\q|{\xi },\q{%
\rangle }_{e(o)}, \qquad {}_e{\langle }\xi ,\q|\xi ,\q{\rangle }%
_o=0.  \label{e14}
\end{equation}
The solutions to equation (14) are 
\begin{eqnarray}
|{\xi },\q{\rangle }_e&=&N_{\sq}^e{\ \sum_{p=\max(0,-\q/2%
)}^\infty }\frac{{\xi }^{2p+\min(0,\sq)}}{{\{[2p]![2p+\q]!\}}^{1/2}}%
|2p+\q,2p{\rangle } \nonumber \\ 
&=&\left\{ 
\begin{array}{ll}
N_{\sq}^e{\sum\limits_{n=0}^\infty }\frac{{\xi }^{2n}}{{\{[2n]![2n+\sq%
]!\}}^{1/2}}|2n+\q,2n{\rangle }, & \q\geq 0, \\[0.2cm] 
N_{\sq}^e{\sum\limits_{n=0}^\infty }\frac{{\xi }^{2n}}{{\{[2n]![2n-\sq%
]!\}}^{1/2}}|2n,2n-\q{\rangle }, & \q\leq 0,
\end{array}
\right.  \label{e15}
\end{eqnarray}
\begin{eqnarray}
|{\xi },\q{\rangle }_o&=&N_{\sq}^o{\ \sum_{p=\max(0,-\q/2%
)}^\infty }\frac{{\xi }^{2p+1+\min(0,\sq)}}{{\{[2p+1]![2p+1+\q]!\}}%
^{1/2}}|2p+1+\q,2p+1{\rangle } \nonumber \\
&=&\left\{ 
\begin{array}{ll}
N_{\sq}^o{\sum\limits_{n=0}^\infty }\frac{{\xi }^{2n+1}}{{\{[2n+1]![2n+1+%
\sq]!\}}^{1/2}}|2n+1+\q,2n+1{\rangle }, & \q\geq 0,
\\ 
N_{\sq}^o{\sum\limits_{n=0}^\infty }\frac{{\xi }^{2n+1}}{{\{[2n+1]![2n+1-%
\sq]!\}}^{1/2}}|2n+1,2n+1-\q{\rangle }, & \q\leq 0,
\end{array}
\right.  \label{e16}
\end{eqnarray}
where the normalization factors $N_{\sq}^{e,o}$ are given by 
\begin{equation}
N_{\sq}^e{\equiv }N_{\sq}^e({|\xi |}^2)={\ \left\{
\sum\limits_{n=0}^\infty \frac{{({|\xi |}^2)}^{2n}}{[2n]![2n+|\q|]!}%
\right\} }^{-{1/2}},  \label{e17}
\end{equation}
\begin{equation}
N_{\sq}^o{\equiv }N_{\sq}^o({|\xi |}^2)={\ \left\{
\sum\limits_{n=0}^\infty \frac{{({|\xi |}^2)}^{2n+1}}{[2n+1]![2n+1+|\q|]!}%
\right\} }^{-{1/2}}.  \label{e18}
\end{equation}
In the limit $q{\rightarrow}1$, they
reduce to the usual even and odd charge coherent states constructed in Ref.~[34]. Note
that $|{\xi },\q{\rangle }_{e(o)}$ are not eigenstates of $a_1a_2$.

{}From (15) and (16), it follows that 
\begin{eqnarray}
&&{}_{e(o)}{\langle }\xi ,\q|{\xi }^{\prime },\q^{\prime }{\rangle }%
_{e(o)}=N_{\sq}^{e(o)}({|\xi |}^2)N_{\sq}^{e(o)}({|{\xi }^{\prime }|}^2){%
\left[ N_{\sq}^{e(o)}({\xi }^{*}{\xi }^{\prime })\right] }^{-2}{\delta }_{%
\q,\q'}, \\
&&{}_e{\langle }\xi ,\q|{\xi }^{\prime },\q^{\prime }{\rangle }_o=0.
\end{eqnarray}
This further indicates that the even (odd) $q$-deformed charge coherent
states are orthogonal with respect to the charge number $\q$ and that
arbitrary even and odd $q$-deformed charge coherent states are orthogonal to
each other. However, these even (odd) states are nonorthogonal with respect
to the parameter $\xi$.

{}For the mean values of the operators $N_1$ and $N_2$, there exists the
relation
\begin{equation}
{}_{e(o)}{\langle }\xi ,\q|N_1|{\xi },\q{\rangle }_{e(o)}=\q + {}_{e(o)}%
{\langle }\xi ,\q|N_2|\xi,\q{\rangle }_{e(o)}.  \label{e21}
\end{equation}

In terms of the even and odd $q$-deformed charge coherent states, the $q$%
-deformed charge coherent states can be expanded as 
\begin{equation}
|\xi ,\q{\rangle }=N_{\sq}\left[ {(N_{\sq}^e)}^{-1}|\xi ,\q{%
\rangle }_e+{(N_{\sq}^o)}^{-1}|\xi ,\q{\rangle }_o\right], \label{e22}
\end{equation}
where the normalization factors are such that 
\begin{equation}
N_{\sq}^{-2}={(N_{\sq}^e)}^{-2}+{(N_{\sq}^o)}^{-2}.  \label{e23}
\end{equation}


\section*{\boldmath 3. Completeness of even and odd $q$-deformed charge coherent
states }

\mbox{}\hspace{6mm}Let us begin with some $q$-deformed formulas which are
useful in the proof of completeness of the even and odd $q$-deformed charge
coherent states. The $q$-deformed Bessel function of (integer) order $\nu $ may be
defined by~[55] 
\begin{equation}
J_\nu (q,x)=\sum\limits_{k=0}^\infty \frac{{(-1)}^k}{[k]![\nu
+k]!}{\left( \frac x{\sqrt{q}[2]_{\sqrt{q}}}\right) }^{\nu +2k},  \label{e24}
\end{equation}
where $[n]_{\sqrt{q}}$ is defined as in equation (5) except for replacing
$q$ by $\sqrt{q}$. An integral representation of the
$q$-deformed modified Bessel function of order $\nu$ is given by~[56] 
\begin{equation}
K_\nu (q,x)=\frac 1{[2]_{\sqrt{q}}}{\left( \frac x{[2]_{\sqrt{q}}}\right) }%
^\nu \int\limits_0^\infty d_qt\frac 1{t^{\nu +1}}e_q(-t)e_q\left( -\frac{x^2%
}{{([2]_{\sqrt{q}})}^2t}\right),  \label{e25}
\end{equation}
where $d_qt$ is a standard $q$-integration [44,57,58], and $e_q(x)$ is a $q$%
-exponential function~[44] 
\begin{equation}
e_q(x)=\left\{ 
\begin{array}{ll}
{\sum\limits_{n=0}^\infty }\frac{x^n}{[n]!}, &{\rm for\ }x>-{\zeta }, \\[0.2cm] 
0, & {\rm otherwise},
\end{array}
\right.  \label{e26}
\end{equation}
with $-\zeta $ being the largest zero of $e_q(x)$. Then, it follows that [56] 
\begin{equation}
\int\limits_0^\infty d_{\sqrt{q}}u\,u^{2p+{\nu }+1}K_\nu (q,[2]_{\sqrt{q}}u)=%
\frac{[\nu +p]![p]!}{{([2]_{\sqrt{q}})}^2}.  \label{e27}
\end{equation}

We now prove that the even and odd $q$-deformed charge coherent states
form an (over)complete set, that is to say 
\begin{equation}
\sum\limits_{\sq={-\infty }}^\infty \int \frac{d_{q}^2{\xi }}\pi {\phi }_{%
\sq}(\xi )N_{\sq}^2\left[ {(N_{\sq}^e)}^{-2}|\xi ,\q\rangle _e\,{}_e{%
\langle }\xi ,\q|+{(N_{\sq}^o)}^{-2}|\xi ,\q\rangle _o\,{}_o{\langle }%
\xi ,\q|\right] \equiv \sum\limits_{\sq={-\infty }}^\infty I_{\sq}=I,
\label{e28}
\end{equation}
where 
\begin{equation}
d_{q}^2{\xi }=|\xi|d_{\sqrt{q}}|\xi|d{\theta}, \qquad {\xi}%
=|\xi|e^{i\theta},  \label{e29}
\end{equation}
and 
\begin{equation}
{\phi }_{\sq}(\xi )= \frac{([2]_{\sqrt{q}})^2 }{2} (-{\rm i})^{\sq}J_{\sq%
}(q, {\rm i}\sqrt{q}[2]_{\sqrt{q}}|\xi|) K_{\sq}(q,[2]_{\sqrt{q}}|\xi|).
\label{e30}
\end{equation}
Note that the integral over $\theta$ is a standard integration while that over $%
|\xi|$ is a $q$-integration. 

In fact, for $\q\geq 0$, we have 
\begin{eqnarray}
I_{\q} &=&\int \frac{d_q^2{\xi }}\pi {\phi }_{\q}(\xi )N_{\sq%
}^2\sum\limits_{j=0}^1\sum\limits_{n,m}^{}\frac{{\xi }^{2n+j}{{\xi }^{*}}%
^{2m+j}|2n+j+\q,2n+j{\rangle }{\langle }2m+j+\q,2m+j|}{{\{[2n+j]![2n+j+%
\q]![2m+j]![2m+j+\q]!\}}^{1/2}}  \nonumber \\
&=&\int\limits_0^\infty \frac{d_{\sqrt{q}}|\xi |}\pi \frac{{([2]_{\sqrt{q}})}%
^2 }{2} {|\xi |}^{\sq+1} K_{\sq}(q,[2]_{\sqrt{q}}|\xi|)
\sum\limits_{j=0}^1\sum\limits_{n,m}^{}{|\xi |}^{2(n+m+j)}\int\limits_{-\pi
}^\pi d{\theta }e^{2{\rm i}(n-m){\theta }}  \nonumber \\
&&\mbox{} {\times }\frac{|2n+j+\q,2n+j{\rangle }{\langle }2m+j+\q,2m+j|}{{%
\{[2n+j]![2n+j+\q]![2m+j]![2m+j+\q]!\}}^{1/2}}  \nonumber \\
&=& \int\limits_0^\infty {d_{\sqrt{q}}|\xi |} {([2]_{\sqrt{q}})}^2 {|\xi |}^{%
\q+1} K_{\sq}(q,[2]_{\sqrt{q}}|\xi|)
\nonumber \\
&& \mbox{} \times
\sum\limits_{j=0}^1\sum\limits_{n=0}^\infty \frac{{\ ({|\xi |}^2)}%
^{2n+j}|2n+j+\q,2n+j{\rangle }{\langle }2n+j+\q,2n+j|}{[2n+j]![2n+j+\q]!} 
\nonumber \\ 
&=&\sum\limits_{n=0}^\infty \frac {{([2]_{\sqrt{q}})}^2 } {[n]![n+\q]!}|n+%
\q,n{\rangle }{\langle }n+\q,n|\int\limits_0^\infty {d_{\sqrt{q}}|\xi |%
}{|\xi |}^{2n+\q+1} K_{\q}(q,[2]_{\sqrt{q}}|\xi|)  \nonumber \\
&=&\sum\limits_{n=0}^\infty |n+\q,n{\rangle }{\langle }n+\q,n|.
\end{eqnarray}
Similarly, for $\q\leq 0$, we get 
\begin{equation}
I_{\sq}=\sum\limits_{n=0}^\infty |n,n-\q{\rangle }{\langle }n,n-\q|.
\label{e32}
\end{equation}
Consequently, we derive 
\begin{eqnarray}
\sum\limits_{\sq=-\infty }^\infty I_{\sq} &=&\sum\limits_{n=0}^\infty
\left( \sum\limits_{\sq=-\infty }^{-1}|n,n-\q{\rangle }{\langle
}n,n-\q|+\sum\limits_{\sq=0}^\infty |n+\q,n{\rangle }{\langle }n+\q%
,n|\right)  \nonumber \\ 
&=&\sum\limits_{m=0}^\infty \sum\limits_{n=0}^\infty
|m,n{\rangle }{\langle } m,n|=I.
\end{eqnarray}
Hence, the even and odd $q$-deformed charge coherent states are qualified to
make up an overcomplete representation. It should be mentioned that $I_{\sq}$
represents the resolution of unity in the subspace where $Q=\q$.


\section*{\boldmath 4. Generation of even and odd $q$-deformed charge coherent states}

\mbox{}\hspace{6mm}By means of (12), (15) and (16), we find 
\begin{eqnarray}
|\xi ,\q\rangle_e &=&{\frac 12}\frac{N_{\sq}^e}{N_{\sq}}(|\xi ,\q\rangle +|-\xi
,\q{\rangle }), \\ 
|\xi ,\q{\rangle }_o &=&{\frac 12}\frac{N_{\sq}^o}{N_{\sq}}(|\xi ,\q\rangle -|-\xi
,\q{\rangle }).
\end{eqnarray}
This shows that the even (odd) $q$-deformed charge coherent states can be
generated by the symmetric (antisymmetric) combination of $q$-deformed
charge coherent states as the charge is conserved. This is similar to the case
of even (odd) $q$-deformed coherent states, which are combinations of $q$%
-deformed coherent states, namely, 
\begin{eqnarray}
|\xi {\rangle }_e &=&{\frac 12}\frac{N^e}N(|\xi {\rangle }+|-\xi {\rangle }%
), \\
|\xi {\rangle }_o &=&{\frac 12}\frac{N^o}N(|\xi {\rangle }-|-\xi {\rangle }%
),
\end{eqnarray}
where 
\begin{equation}
\label{e38}
|\xi {\rangle }=N\sum\limits_{n=0}^\infty \frac{{\xi }^n}{\sqrt{[n]!}}|n{%
\rangle },
\end{equation}
\begin{equation}
\label{e39}
N=e_q^{-{1/2}}({|\xi |}^2), \quad
N^e{\equiv }N^e({|\xi |}%
^2)=\cosh_q^{-{1/2}}{|\xi |}^2, \quad
N^o{\equiv }N^o({|\xi |}%
^2)=\sinh_q^{-{1/2}}{|\xi |}^2,
\end{equation}
with
\begin{eqnarray}
&&\cosh_qx=\sum\limits_{n=0}^\infty\frac{x^{2n}}{[2n]!}, \\
&&\sinh_qx=\sum\limits_{n=0}^\infty\frac{x^{2n+1}}{[2n+1]!}.
\end{eqnarray}

The even (odd) $q$-deformed charge coherent states can also be obtained from
the states $(36)-(38)$ according to the following expression 
\begin{eqnarray}
\lefteqn{|\xi ,\q{\rangle }_{e(o)}} \nonumber \\
&=&\left\{ 
\begin{array}{ll}
N_{\sq}^{e(o)}e_q^{{1/2}} ({|{\xi }_1|}^2) {\left[ N^{e(o)}({|{\xi }_2|}%
^2)\right] }^{-1}{\xi _1}^{-\sq}\int\limits_{-\pi }^\pi \frac{d{\alpha }}{%
2\pi }e^{+{\rm i}\sq\alpha }|e^{-{\rm i}\alpha }{\xi }_1{\rangle }{\otimes
}|e^{{\rm i}\alpha }{%
\xi }_2{\rangle }_{e(o)}, & \q\geq 0, \\[0.2cm] 
N_{\sq}^{e(o)} e_q^{{1/2}} ({|{\xi }_1|}^2) {\left[ N^{e(o)}({|{\xi }_2|}%
^2)\right] }^{-1}{\xi _1}^{+\sq}\int\limits_{-\pi }^\pi \frac{d{\alpha }}{%
2\pi }e^{-{\rm i}\sq\alpha }|e^{{\rm i}\alpha }{\xi }_2{\rangle }_{e(o)}{\otimes }%
|e^{-{\rm i}\alpha }{\xi }_1{\rangle }, & \q\leq 0,
\end{array}
\right.  \label{e42}
\end{eqnarray}
where $\xi ={\xi }_1{\xi }_2$. Such a representation is very useful since
the properties of $q$-deformed coherent states and even (odd) $q$-deformed
coherent states can now be employed in a study of the properties of even
(odd) $q$-deformed charge coherent states. The expression for the latter given in (42)
has a very simple group-theoretical interpretation: in (42) one suitably averages over the
U(1)-group (caused by the charge operator $Q$) action on the product of $q$%
-deformed coherent states and even (odd) $q$-deformed coherent states, which
then projects out the $Q=\q$ charge subspace contribution.

It is easy to see that in the limit $q{\rightarrow}1$, the above discussion
gives back the corresponding results for the usual even and odd charge
coherent states obtained in Ref.~[34].


\section*{\boldmath 5. $D$-algebra realization of SU$_q(1,1)$ generators}

\mbox{}\hspace{6mm}As is well known, the coherent state $D$-algebra [6,59] is a
mapping of quantum observables onto a differential form that acts on the parameter space
of coherent states, and has a beautiful application in the reformulation of the entire laser
theory in terms of $C$-number differential equations [60]. We shall construct the
$D$-algebra realization of the $q$-deformed SU$_q(1,1)$ generators corresponding to
the unnormalized even and odd $q$-deformed charge coherent states, defined by 
\begin{equation}
||\q{\rangle }_{e(o)}{\equiv }||\xi ,\q{\rangle }_{e(o)}={\left[ N_{\sq}^{e(o)}\right]
}^{-1}|\xi ,\q{\rangle }_{e(o)}.  \label{e43}
\end{equation}

The action of the operators $a_i$, $a_i^{+}$ and $N_i$ on the column vector 
composed of $||\q{\rangle }_e$ and $||\q{\rangle }_o$ can be written
in the matrix form: 

\begin{equation}
\begin{array}{ll}
{\rm Positive\ } Q & {\rm Negative\ } Q \\[0.4cm]
a_1\left[ 
\begin{array}{l}
||\q{\rangle }_e \\ 
||\q{\rangle }_o
\end{array}
\right] =\left[ 
\begin{array}{l}
||\q-1{\rangle }_e \\ 
||\q-1{\rangle }_o
\end{array}
\right], & a_1\left[ 
\begin{array}{l}
||\q{\rangle }_e \\ 
||\q{\rangle }_o
\end{array}
\right] ={\xi }\left[ 
\begin{array}{lr}
0 & 1 \\ 
1 & 0
\end{array}
\right] \left[ 
\begin{array}{l}
||\q-1{\rangle }_e \\ 
||\q-1{\rangle }_o
\end{array}
\right],  \\[0.4cm]
a_2\left[ 
\begin{array}{l}
||\q{\rangle }_e \\ 
||\q{\rangle }_o
\end{array}
\right] ={\xi }\left[ 
\begin{array}{lr}
0 & 1 \\ 
1 & 0
\end{array}
\right] \left[ 
\begin{array}{l}
||\q+1{\rangle }_e \\ 
||\q+1{\rangle }_o
\end{array}
\right], & a_2\left[ 
\begin{array}{l}
||\q{\rangle }_e \\ 
||\q{\rangle }_o
\end{array}
\right] =\left[ 
\begin{array}{l}
||\q+1{\rangle }_e \\ 
||\q+1{\rangle }_o
\end{array}
\right], \\[0.4cm]
a_1^{+}\left[ 
\begin{array}{l}
||\q{\rangle }_e \\ 
||\q{\rangle }_o
\end{array}
\right] ={\xi }^{-\sq}\frac d{d_q\xi }{\xi}^{\sq+1}\left[ 
\begin{array}{l}
||\q+1{\rangle }_e \\ 
||\q+1{\rangle }_o
\end{array}
\right], & a_1^{+}\left[ 
\begin{array}{l}
||\q{\rangle }_e \\ 
||\q{\rangle }_o
\end{array}
\right] = \frac d{d_q\xi } \left[ 
\begin{array}{lr}
0 & 1 \\ 
1 & 0
\end{array}
\right] \left[ 
\begin{array}{l}
||\q+1{\rangle }_e \\ 
||\q+1{\rangle }_o
\end{array}
\right],  \\[0.4cm]
a_2^{+}\left[ 
\begin{array}{l}
||\q{\rangle }_e \\ 
||\q{\rangle }_o
\end{array}
\right] =\frac d{d_q\xi }\left[ 
\begin{array}{lr}
0 & 1 \\ 
1 & 0
\end{array}
\right] \left[ 
\begin{array}{l}
||\q-1{\rangle }_e \\ 
||\q-1{\rangle }_o
\end{array}
\right], & a_2^{+}\left[ 
\begin{array}{l}
||\q{\rangle }_e \\ 
||\q{\rangle }_o
\end{array}
\right] = {\xi }^{\sq}\frac d{d_q\xi }{\xi}^{-\sq+1} \left[ 
\begin{array}{l}
||\q-1{\rangle }_e \\ 
||\q-1{\rangle }_o
\end{array}
\right], \\[0.4cm]
N_1\left[ 
\begin{array}{l}
||\q{\rangle }_e \\ 
||\q{\rangle }_o
\end{array}
\right] = \left({\xi }\frac d{d\xi }+\q \right) \left[ 
\begin{array}{l}
||\q{\rangle }_e \\ 
||\q{\rangle }_o
\end{array}
\right], & N_1\left[ 
\begin{array}{l}
||\q{\rangle }_e \\ 
||\q{\rangle }_o
\end{array}
\right] = \xi\frac d{d\xi } \left[ 
\begin{array}{l}
||\q{\rangle }_e \\ 
||\q{\rangle }_o
\end{array}
\right], \\[0.4cm]
N_2\left[ 
\begin{array}{l}
||\q{\rangle }_e \\ 
||\q{\rangle }_o
\end{array}
\right] = {\xi }\frac d{d\xi } \left[ 
\begin{array}{l}
||\q{\rangle }_e \\ 
||\q{\rangle }_o
\end{array}
\right], & N_2\left[ 
\begin{array}{l}
||\q{\rangle }_e \\ 
||\q{\rangle }_o
\end{array}
\right] = \left(\xi\frac d{d\xi }-\b{q}\right) \left[ 
\begin{array}{l}
||\q{\rangle }_e \\ 
||\q{\rangle }_o
\end{array}
\right],  \label{e49}
\end{array}
\end{equation}
where $d/d\xi$ is a standard differential operator, whereas
$d/d_q\xi$ is a $q$-differential one [39,44,58], defined by 
\begin{equation}  \label{e50}
\frac{d}{d_q\xi}f(x)=\frac{f(qx)-f(q^{-1}x)}{qx-q^{-1}x }.
\end{equation}
The $q$-deformed SU$_q(1,1)$ algebra consists of three generators $K_0$, $%
K_{+}$, and $K_{-}$, satisfying the commutation relations 
\begin{equation}
\lbrack K_{+},K_{-}]=-[2K_0], \qquad [K_0,K_{\pm }]={\pm }K_{\pm },
\label{e51}
\end{equation}
and is realized in terms of the two-mode $q$-boson operators as 
\begin{equation}
K_{-}=a_1a_2, \qquad K_{+}=a_1^{+}a_2^{+}, \qquad K_0={\frac
12}(N_1+N_2+1).  \label{e52}
\end{equation}
Actually, the even and odd $q$-deformed charge coherent states are also the
eigenstates of the square of $K_{-}$.

The $D$-algebra of the SU$_q(1,1)$ generators $A$ may be defined for the action on
the ket coherent states (43) or for that on the corresponding bras as
\begin{eqnarray}
A\left[ 
\begin{array}{l}
||\q{\rangle }_e \\ 
||\q{\rangle }_o
\end{array}
\right] &=&D^k(A)\left[ 
\begin{array}{l}
||\q{\rangle }_e \\ 
||\q{\rangle }_o
\end{array}
\right], \label{e53}\\
\left[ 
\begin{array}{l}
{}_e{\langle }\q|| \\ 
{}_o{\langle }\q||
\end{array}
\right] A &=&D^b(A)\left[ 
\begin{array}{l}
{}_e{\langle }\q|| \\ 
{}_o{\langle }\q||
\end{array}
\right],
\end{eqnarray}
respectively.
Using (\ref{e49}) and (\ref{e52}), we get for the former 
\begin{eqnarray}
D^k(K_{-})& = &{\xi }\left[ 
\begin{array}{lr}
0 & 1 \\ 
1 & 0
\end{array}
\right], \label{e55} \\
D^k(K_{+}) & = & {\xi }^{-|\sq|}\frac d{d_q\xi }{\xi}^{|\sq|+1}\frac d{d_q\xi } \left[ 
\begin{array}{lr}
0 & 1 \\ 
1 & 0
\end{array}
\right], \\
D^k(K_0) & = &\frac 12 \left(2{\xi }\frac d{d\xi }+|\q|+1\right)I,
\end{eqnarray}
while the latter can be obtained from the adjoint relation 
\begin{equation}
D^b(A)={\left[ D^k(A^{+})\right] }^{*}.  \label{e58}
\end{equation}
Thus, the $D$-algebra of the SU$_q(1,1)$ generators corresponding to the unnormalized
even and odd $q$-deformed charge coherent states has been realized in a
$q$-differential-operator matrix form.

{}From (\ref{e53}) and (\ref{e55}), we clearly see that the unnormalized even
and odd
$q$-deformed charge coherent states can be transformed into each other by the
action of the operator $a_1a_2$. Actually, $a_1a_2$ plays the role of a
connecting operator between the two kinds of states.

It is easy to check that in the limit $q{\rightarrow}1$, the above
discussion gives back that carried out in Ref.~[34] for the usual even and odd charge
coherent states.


\section*{\boldmath 6. Nonclassical properties of even and odd $q$-deformed charge
coherent states}

\mbox{}\hspace{6mm}In this section, we will study some nonclassical properties of the
even and odd $q$-deformed charge coherent states, such as 
SU$_q(1,1)$ squeezing, single- or two-mode $q$-squeezing, and two-mode $q$%
-antibunching. 


\subsection*{\boldmath 6.1. SU$_q(1,1)$ squeezing}

\mbox{}\hspace{6mm}In analogy with the definition of SU(1,1) squeezing [61],
we introduce  SU$_q(1,1)$ squeezing in terms of the Hermitian
$q$-deformed quadrature operators 

\begin{equation}
X_1=\frac{K_{+}+K_{-}}2, \qquad X_2=\frac{{\rm i}(K_{+}-K_{-})}2,  \label{e59}
\end{equation}
which satisfy the commutation relation 
\begin{equation}
\lbrack X_1,X_2]=\frac{\rm i}{2}[2K_0]  \label{e60}
\end{equation}
and the uncertainty relation 
\begin{equation}
{\langle }{({\Delta }X_1)}^2{\rangle }{\langle }{({\Delta }X_2)}^2{\rangle }{%
\geq }\frac 1{16}{|{\langle }[2K_0]{\rangle }|}^2.  \label{e61}
\end{equation}
A state is said to be SU$_q(1,1)$ squeezed if 
\begin{equation}
{\langle }{({\Delta }X_i)}^2{\rangle }<\frac 14{|{\langle }[2K_0]{\rangle }|}
\qquad (i=1 {\rm \ or\ }2).  \label{e62}
\end{equation}

Let us now calculate the fluctuations (variances) of $X_1$ and $X_2$ with
respect to the even and odd $q$-deformed charge coherent states. Using 
(\ref{e52}) -- (\ref{e55}) and (\ref{e58}), we get 
\begin{eqnarray}
{}_e{\langle }{\xi },\q|K_{+}K_{-}|\xi ,\q{\rangle }_e & = &{|\xi |}^2%
\, \overline{\tanh}_{\sq}{|\xi |}^2, \\
{}_o{\langle }{\xi },\q|K_{+}K_{-}|\xi ,\q{\rangle }_o & = &{|\xi |}^2%
\, \overline{\coth}_{\sq}{|\xi |}^2,
\end{eqnarray}
where 
\begin{equation}
\overline{\tanh}_{\sq}{|\xi |}^2 \equiv \frac{\overline{\sinh}_{\sq}{|\xi |}^2}{%
\overline{\cosh}_{\sq}{|\xi |}^2}, \qquad \overline{\coth}_{\sq}{|\xi |}%
^2 \equiv \frac 1{\overline{\tanh}_{\sq}{|\xi |}^2},  \label{e65}
\end{equation}
with 
\begin{equation}
\overline{\sinh}_{\sq}x \equiv {\left[ N_{\sq}^o(x)\right] }^{-2}, \qquad
\overline{\cosh}_{\sq}x \equiv {\left[ N_{\sq}^e(x)\right] }^{-2}.  \label{e66}
\end{equation}
Furthermore,
\begin{equation}
{}_{e(o)}{\langle }K_{+}{\rangle }_{e(o)}= {}_{e(o)}{\langle }K_{-}{\rangle }%
_{e(o)}=0.  \label{e67}
\end{equation}
Thus, the fluctuations are given by 
\begin{eqnarray}
{}_e{\langle }{({\Delta }X_1)}^2{\rangle }_e &= &{\frac 14}{}_e{\langle }[2K_0]
\rangle_e+\frac 12{|\xi |}^2(\cos2\theta +\overline{\tanh}_{\sq}{|\xi |}^2),
\label{e68}\\
{}_e{\langle }{({\Delta }X_2)}^2{\rangle }_e & = &{\frac 14}{}_e{\langle }[2K_0]{%
\rangle }_e+\frac 12{|\xi |}^2(-\cos2\theta +\overline{\tanh}_{\sq}{|\xi |}%
^2), \\
{}_o{\langle }{({\Delta }X_1)}^2{\rangle }_o & = &{\frac 14}{}_o{\langle }[2K_0]{%
\rangle }_o+\frac 12{|\xi |}^2(\cos2\theta +\overline{\coth}_{\sq}{|\xi |}%
^2), \\
{}_o{\langle }{({\Delta }X_2)}^2{\rangle }_o & = &{\frac 14}{}_o{\langle }[2K_0]{%
\rangle }_o+\frac 12{|\xi |}^2(-\cos2\theta +\overline{\coth}_{\sq}{|\xi |}%
^2). \label{e71}
\end{eqnarray}

According to (\ref{e68}) -- (\ref{e71}), SU$_q(1,1)$ squeezing of the states $|\xi
,\q{\rangle }_{e(o)}$ occurs as long as 
\begin{eqnarray}
&&{\pm }\cos2{\theta }+\overline{\tanh}_{\sq}{|\xi |}^2<0, \label{e72}\\
&&{\pm }\cos2{\theta }+\overline{\coth}_{\sq}{|\xi |}^2<0,  \label{e73}
\end{eqnarray}
for one of the sign choices. Choose ${\theta }$ to be $\frac \pi 2$ or $0$, 
so that ${\pm }\cos2\theta =-1$. When $|\xi |\leq 1$, $\overline{\tanh}_{\sq}{%
|\xi |}^2<1$. Thus, condition (\ref{e72}) can be satisfied for $|\xi |\leq 1$.
Using (\ref{e17}), (\ref{e18}), (\ref{e23}), (\ref{e65}), and (\ref{e66}), we can prove
that 
\begin{equation}
\overline{\coth_{\sq}} |\xi|^2 = 1 + \frac{1}{|\xi|^{|\sq|}
\overline{\sinh}_{\sq}|\xi|^2} J_{|\sq|}\left(q, \sqrt{q} [2]_{\sqrt{q}} |\xi|\right).
\label{e74}
\end{equation}
We know that $\overline{\sinh}_{\sq}|\xi|^2\geq 0$. Obviously,
$\overline{\coth_{\sq}} |\xi|^2<1$ if
$J_{|\sq|}\left(q, \sqrt{q} [2]_{\sqrt{q}} |\xi|\right)<0$.
In fact, for arbitrary
fixed values of $\q$ and $q$, there surely exists some range of $|\xi|$
values such that $J_{|\sq|}\left(q, \sqrt{q} [2]_{\sqrt{q}} |\xi|\right)<0$.
For instance, numerical calculations show that, for $\q=0$ and $q=0.2$,
$J_0\left(0.2, \sqrt{0.2} [2]_{\sqrt{0.2}} |\xi|\right)<0$
when $1.020\leq |\xi |\leq5.208 $;
for $\q={\pm }1$ and $q=0.5$,
$J_1\left(0.5, \sqrt{0.5} [2]_{\sqrt{0.5}} |\xi|\right)<0$
when $1.808\leq |\xi |\leq 3.770$;
for $\q={\pm }2$ and $q=0.9$,
$J_2\left(0.9, \sqrt{0.9} [2]_{\sqrt{0.9}} |\xi|\right)<0$
when $2.560\leq |\xi |\leq 4.166$. 
Thus, for arbitrary fixed values of $\q$ and $q$, condition (\ref{e73})
can also be satisfied over some limited range of $|\xi |$ values. 

{}From the above discussion, it is apparent that both even and odd $q$%
-deformed charge coherent states can exhibit SU$_q(1,1)$ squeezing.

It is easy to verify that the $q$-deformed charge coherent states satisfy
the equality in (\ref{e61}) and that
${\langle}{({\Delta}X_{1})}^2{\rangle}={%
\langle}{({\Delta}X_{2})}^2{\rangle}$. 
Therefore, the $q$-deformed charge coherent states, contrary to the even and
odd ones, are not SU$_q(1,1)$ squeezed.


\subsection*{\boldmath 6.2. Single-mode $q$-squeezing}

\mbox{}\hspace{6mm}In analogy with the definition of single-mode squeezing
[27], we introduce single-mode $q$-squeezing in terms of the Hermitian
$q$-deformed quadrature operators for the individual modes 
\begin{eqnarray}
Y_1 &=&\frac{a_1^{+}+a_1}2,\qquad Y_2=\frac{{\rm i}(a_1^{+}-a_1)}2,
\nonumber \\
Z_1 &=&\frac{a_2^{+}+a_2}2,\qquad Z_2=\frac{{\rm i}(a_2^{+}-a_2)}2,
\end{eqnarray}
which satisfy the commutation relations 
\begin{equation}
\lbrack Y_1,Y_2]=\frac {\rm i}2[a_1,a_1^+],\qquad [Z_1,Z_2]=\frac
{\rm i}2[a_2,a_2^+],  \label{e76}
\end{equation}
and the uncertainty relations 
\begin{equation}
{\langle }{({\Delta }Y_1)}^2{\rangle }{\langle }{({\Delta }Y_2)}^2{\rangle }{%
\geq }\frac 1{16} {|{\langle }[a_1,a_1^+]{\rangle }|}^2,\qquad {%
\langle }{({\Delta }Z_1)}^2{\rangle }{\langle }{({\Delta }Z_2)}^2{\rangle }{%
\geq }\frac 1{16}{|{\langle }[a_2,a_2^+]{\rangle }|}^2.  \label{e77}
\end{equation}
A state is said to be single-mode $q$-squeezed if 
\begin{equation}
{\langle }{({\Delta }Y_i)}^2{\rangle }<\frac 14 |{\langle }[a_1,a_1^+]{%
\rangle }|,\qquad {\langle }{({\Delta }Z_i)}^2{\rangle }<\frac 14 |{%
\langle }[a_2,a_2^+]{\rangle }|~\hspace{5mm}(i=1{\rm \ or\ }2).  \label{e78}
\end{equation}

{}For the even and odd $q$-deformed charge coherent states, it always follows
that 
\begin{equation}  \label{e79}
{}_{e(o)}{\langle}a_1{\rangle}_{e(o)}={}_{e(o)}{\langle}a_2{\rangle}_{e(o)}=
{}_{e(o)}{%
\langle}a_1^2{\rangle}_{e(o)}={}_{e(o)}{\langle}a_2^2{\rangle}_{e(o)}= {}_{e(o)}{%
\langle}a_1^+a_2{\rangle}_{e(o)}=0.
\end{equation}
Thus, the fluctuations are given by
\begin{eqnarray}
{}_{e(o)}{\langle}{({\Delta}Y_{1})}^2{\rangle}_{e(o)}={}_{e(o)}{\langle}{({\Delta%
}Y_{2})}^2{\rangle}_{e(o)}&=& \frac{1}{4}\left( {}_{e(o)}{\langle}[a_1,a_1^+]{%
\rangle}_{e(o)}+2\, {}_{e(o)}{\langle}a_1^+a_1{\rangle}_{e(o)}\right)  \nonumber
\\
&>&\frac{1}{4}{}_{e(o)}{\langle}[a_1,a_1^+]{\rangle}_{e(o)},
\end{eqnarray}
\begin{eqnarray}
{}_{e(o)}{\langle}{({\Delta}Z_{1})}^2{\rangle}_{e(o)}={}_{e(o)}{\langle}{({\Delta%
}Z_{2})}^2{\rangle}_{e(o)}&=& \frac{1}{4}\left( {}_{e(o)}{\langle}[a_2,a_2^+]{%
\rangle}_{e(o)}+2\, {}_{e(o)}{\langle}a_2^+a_2{\rangle}_{e(o)}\right)  \nonumber
\\
&>&\frac{1}{4}{}_{e(o)}{\langle}[a_2,a_2^+]{\rangle}_{e(o)}.
\end{eqnarray}
This shows that there is no single-mode $q$-squeezing in both even and odd $q$%
-deformed charge coherent states. The same situation occurs for the $q$%
-deformed charge coherent states [62].


\subsection*{\boldmath 6.3. Two-mode $q$-squeezing}

\mbox{}\hspace{6mm}In analogy with the definition of two-mode squeezing
[63], we introduce two-mode $q$-squeezing in terms of the Hermitian
$q$-deformed quadrature operators for the two modes 
\begin{equation}  \label{e82}
W_1=\frac{Y_1+Z_1}{\sqrt{2}}=\frac{1}{\sqrt{8}}(a_1^{+}+a_2^{+}+a_1+a_2),\quad
W_2=\frac{Y_2+Z_2}{\sqrt{2}}=\frac{\rm i}{\sqrt{8}}(a_1^{+}+a_2^{+}-a_1-a_2),
\end{equation}
which satisfy the commutation relation 
\begin{equation}  \label{e83}
[W_1,W_2]=\frac{1}{4}{\rm i}\left\{[a_1,a_1^+]+[a_2,a_2^+]\right\}
\end{equation}
and the uncertainty relation 
\begin{equation}  \label{e84}
{\langle}{({\Delta}W_{1})}^2{\rangle}{\langle}{({\Delta}W_{2})}^2{\rangle}{%
\geq}\frac{1}{64} {|{\langle }[a_1,a_1^+]{\rangle }+{\langle }[a_2,a_2^+]{%
\rangle } |}^2.
\end{equation}
A state is said to be two-mode $q$-squeezed if 
\begin{equation}  \label{e85}
{\langle}{({\Delta}W_{i})}^2{\rangle}<\frac{1}{8} {|{\langle }[a_1,a_1^+]{%
\rangle }+{\langle }[a_2,a_2^+]{\rangle } |} ~ \hspace{8mm}(i=1 {\rm \ or\ }2).
\end{equation}

{}For the even and odd $q$-deformed charge coherent states, the fluctuations
are given by 
\begin{eqnarray}
{}_{e(o)}{\langle}{({\Delta }W_1)}^2{\rangle}_{e(o)}&=&{}_{e(o)}{\langle}{({%
\Delta }W_2)}^2{\rangle}_{e(o)}=\frac 12\left({}_{e(o)}{\langle}{({\Delta }Y_1)%
}^2{\rangle}_{e(o)}+{}_{e(o)}{\langle}{({\Delta }Z_1)}^2{\rangle}_{e(o)}\right)
\nonumber \\
&=&\frac 12\left({}_{e(o)}{\langle }{({\Delta }Y_2)}^2{\rangle }_{e(o)}+{}_{e(o)}%
{\langle }{({\Delta }Z_2)}^2{\rangle }_{e(o)}\right)  \nonumber \\
&=&\frac 18 \Bigl( {}_{e(o)}{\langle}[a_1,a_1^+]{\rangle}_{e(o)}+ {}_{e(o)}{%
\langle}[a_2,a_2^+]{\rangle}_{e(o)} +2\,{}_{e(o)}{\langle}a_1^+a_1{\rangle}%
_{e(o)} \nonumber \\
&&\mbox{} +2\,{}_{e(o)}{\langle}a_2^+a_2{\rangle}_{e(o)} \Bigr)  \nonumber\\ 
&>&\frac{1}{8} \left( {}_{e(o)}{\langle}[a_1,a_1^+]{\rangle}_{e(o)}+ {}_{e(o)}{%
\langle}[a_2,a_2^+]{\rangle}_{e(o)} \right).
\end{eqnarray}
This shows there is no two-mode $q$-squeezing in both even and odd $q$%
-deformed charge coherent states. On the contrary, there is such $q$-squeezing in
the $q$-deformed charge coherent states [61].


\subsection*{\boldmath 6.4. Two-mode $q$-antibunching}

\mbox{}\hspace{6mm}In analogy with the definition of two-mode antibunching
[34], we introduce a two-mode $q$-correlation function as 
\begin{equation}
g\equiv \frac{{\langle }{(a_1^{+}a_2^{+})}^2{(a_1a_2)}^2{\rangle }}{{{%
\langle }a_1^{+}a_2^{+}a_1a_2{\rangle }}^2}=\frac{{\langle }:{([N_1][N_2])}%
^2:{\rangle }}{{{\langle }[N_1][N_2]{\rangle }}^2},  \label{e87}
\end{equation}
where $a_i$ and $a_i^{+}$ represent the annihilation and creation operators
of $q$-deformed photons of a deformed light field and $:\,:$ denotes normal
ordering. Physically, $g$ is a measure of $q$-deformed two-photon
correlations in the $q$-deformed two-mode field and is related to the $q$%
-deformed two-photon number distributions. A state is said to be two-mode $%
q$-antibunched if 
\begin{equation}
g<1.  \label{e88}
\end{equation}

{}For the even and odd $q$-deformed charge coherent states, we have 
\begin{eqnarray}
g_e &=&{\overline{\coth}_{\sq}}^2{|\xi |}^2, \\
g_o &=&{\overline{\tanh}_{\sq}}^2{|\xi |}^2.
\end{eqnarray}
{}From the above discussion about the function $\overline{\coth}_{\sq}{|\xi |}%
^2$ $(\overline{\tanh}_{\sq}{|\xi |}^2)$, we see that $g_{e(o)}$ can be less
than $1$ 
over some particular range of $|\xi|$ values. This indicates that for both
even and odd $q$-deformed charge coherent states, two-mode $q$-antibunching
exists. On the contrary, for the $q$-deformed charge coherent states we have $g=1$
so that no two-mode $q$-antibunching exists. 

It can be shown that in the limit $q{\rightarrow }1$, the nonclassical properties of the
usual even and odd charge coherent states, studied in Ref. [34], are retrieved as 
expected.


\section*{7. Summary}

\mbox{}\hspace{6mm}Let us sum up the results obtained in the present Letter:

(1) The even and odd $q$-deformed charge coherent states, defined as the two
orthonormalized eigenstates of both the square of the pair $q$-boson
annihilation operator and the charge operator, have been constructed and their
(over)completeness proved. Such $q$-deformed states become the usual even
and odd charge coherent states in the limit $q{\rightarrow}1$.

(2) The even (odd) $q$-deformed charge coherent states have been shown to be
generated by the symmetric (antisymmetric) combination of $q$-deformed charge
coherent states in the case of conserved charge, or by a suitable average over the $%
U(1)$-group (caused by the charge operator) action on the product of $q$%
-deformed coherent states and even (odd) $q$-deformed coherent states.

(3) The $D$-algebra of the SU$_q(1,1)$ generators corresponding to the even and odd
$q$-deformed charge coherent states has been realized in a $q$-differential
operator matrix form.

(4) Both even and odd $q$-deformed charge coherent states have been shown to exhibit
SU$_q(1,1)$ squeezing and two-mode $q$-antibunching, but neither single-mode
nor two-mode $q$-squeezing.

\section*{Acknowledgments}

\mbox{}\hspace{6mm}X.-M.L.\ is grateful to Professor C. Quesne for warm
hospitality at the Universit\'{e} Libre de Bruxelles. He also acknowledges
financial support by the National Fund for Scientific Research (FNRS), Belgium, as well as
support by the National Natural Science Foundation of China under Grants Nos.
10174007, 10074008 and 60278021. C.Q.\ is a Research Director of the National Fund
for Scientific Research (FNRS), Belgium.   


\end{document}